\documentclass[aps,prb,reprint,showpacs,superscriptaddress,floatfix]{revtex4-1}
\usepackage{bm}
\usepackage{amssymb}
\usepackage{graphicx}
\usepackage{amsmath}
\usepackage{dcolumn}
\usepackage[pdftex]{epsfig}
\usepackage{color}
\usepackage[german,english]{babel}
\usepackage{psfrag}
\usepackage[urlcolor=blue,linkcolor=blue,citecolor=blue,colorlinks=true]{hyperref}
\newcommand{\msu}{\uparrow}			
\newcommand{\msd}{\downarrow}			

\begin{document}

\title{First-order metal-insulator transitions in the extended Hubbard model due to self-consistent screening of the effective interaction}
\author{M. Sch\"uler}
\email{mschueler@itp.uni-bremen.de}
\affiliation{Institut f{\"u}r Theoretische Physik, Universit{\"a}t Bremen, Otto-Hahn-Allee 1, 28359 Bremen, Germany}
\affiliation{Bremen Center for Computational Materials Science, Universit{\"a}t Bremen, Am Fallturm 1a, 28359 Bremen, Germany}
\author{E. G. C. P. van Loon}
\author{M. I. Katsnelson}
\affiliation{Radboud University, Institute for Molecules and Materials, Heyendaalseweg 135, NL-6525 AJ Nijmegen, The Netherlands}
\author{T. O. Wehling}
\affiliation{Institut f{\"u}r Theoretische Physik, Universit{\"a}t Bremen, Otto-Hahn-Allee 1, 28359 Bremen, Germany}
\affiliation{Bremen Center for Computational Materials Science, Universit{\"a}t Bremen, Am Fallturm 1a, 28359 Bremen, Germany}

\date{\today}

\begin{abstract}
While the Hubbard model is the standard model to study Mott metal-insulator transitions, it is still unclear to what extent it can describe metal-insulator transitions in real solids, where nonlocal Coulomb interactions are always present. By using a variational principle, we clarify this issue for short- and long-range nonlocal Coulomb interactions for half-filled systems on bipartite lattices. We find that repulsive nonlocal interactions generally stabilize the Fermi-liquid regime. The metal-insulator phase boundary is shifted to larger interaction strengths to leading order linearly with nonlocal interactions. Importantly, nonlocal interactions can raise the order of the metal-insulator transition. We present a detailed analysis of how the dimension and geometry of the lattice as well as the temperature determine the critical nonlocal interaction leading to a first-order transition: for systems in more than two dimensions with non-zero density of states at the Fermi energy the critical nonlocal interaction is arbitrarily small; otherwise, it is finite.
\end{abstract}

\maketitle
\section{Introduction}
The Hubbard model \cite{Hubbard63,Gutzwiller63,Kanamori63,Hubbard64,Gutzwiller64} is a central model for understanding various aspects of strongly correlated electrons. It incorporates the competition between kinetic and interaction energies in the most basic way and exhibits phenomena such as magnetism and metal-insulator transitions with and without magnetic transitions \cite{slater_magnetic_1951,slater_self-consistent_1974,mottmetal-insulator1990,gebhard_florian_mott_1997}. However, particularly due to neglecting nonlocal interaction, the Hubbard model can be quite far from describing real materials whenever nonlocal interactions are not efficiently screened, e.g., in two-dimensional materials. One example is the plasmon dispersion in metals which differs qualitatively in models with and without nonlocal interactions \cite{hafermann_collective_2014,van_loon_plasmons_2014}. Most obviously, in insulating systems, where screening is by definition incomplete, prominent nonlocal interaction effects should be expected. It is thus unclear whether the Hubbard model can describe the Mott metal-insulator transition (MIT) even qualitatively correctly. 

Indeed, the question about the order of the MIT has been controversial for about five decades \cite{mott_basis_1949,mott_metal-insulator_1968,imada_metal-insulator_1998}. In the Hubbard model with strictly local interaction, the order of the MIT depends on the degree of magnetic frustration in the system.  If magnetic order is fully suppressed, the transition is of first order below a critical temperature $T_c$ as, e.g., dynamical mean-field theory (DMFT) \cite{blumer_mott-hubbard_2003} and related quantum cluster theories \cite{park_cluster_2008,balzer_first-order_2009,merino_pseudogap_2014} have demonstrated. Otherwise, the MIT is accompanied by magnetic (quasi)order and is continuous \cite{santoro_hubbard_1993,ulmke_anderson-hubbard_1995,schafer_fate_2015}. Thus, Hubbard models on bipartite lattices, like the honeycomb, square, diamond, and cubic lattice, as well as higher dimensional generalizations thereof, feature continuous MITs. Due to the various simplifications implied by the Hubbard model, it is, however, unclear how well this picture of the MIT relates to the experimentally realized one. Already in his original work, Mott, for instance, argues that the physically realized MIT should, due to the long-range nature of the Coulomb interaction, be of first order \cite{mott_basis_1949}, which is indeed found experimentally in many transition-metal oxides \cite{morin_oxides_1959,mcwhan_metal-insulator_1970,shivashankar_metalchar22antiferromagnetic-insulator_1983}. In this context, the question of how the Hubbard model's MIT is connected to that of the extended Hubbard model, which includes nonlocal interactions, is highly relevant.

\begin{figure}[b]
\mbox{
\includegraphics[width=0.99\columnwidth]{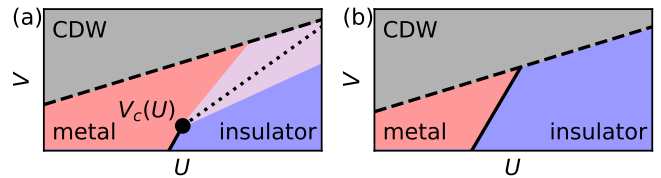}
}
\caption{(Color online) Schematic phase diagram of the extended Hubbard model: (a) The continuous (solid) and first-order (dotted) metal-insulator transition lines touch at $V_c(U)$. A coexistence region surrounds the first-order transition. For large $V/U$ a CDW phase occurs. (b) The CDW phase can mask the first-order MIT. }
\label{fig:schem}
\end{figure}

Here, we show that the MIT in the half-filled extended Hubbard model on bipartite lattices is of first order for nonlocal interactions larger than a critical $V_c$, as depicted schematically in Fig.~\ref{fig:schem}(a). $V_c$ depends on the dimension and lattice topology  and can be even arbitrarily small in cubic systems in $d>2$. The first-order transition can be masked by a charge density wave [CDW; Fig.~\ref{fig:schem}(b)], a situation which we find, e.g., in the honeycomb lattice.

We propose (and later substantiate) the following mechanism for how nonlocal interactions induce a discontinuous MIT: Nonlocal interactions generally decrease correlations in half-filled extended Hubbard models \cite{schuler_optimal_2013,van_loon_capturing_2016}. The amount of decrease, due to different screening \cite{ayral_screening_2013}, is larger in the metallic regime than in the insulating regime. Now consider two systems close to the MIT with initially no nonlocal interactions: one metallic and one insulating. Nonlocal interactions $V$ will push the MIT of the formerly metallic system to larger local interaction $U_c^\text{met.} > U_c^\text{ins.}$ than the formerly insulating system, resulting in a discontinuous (i.e., first-order) MIT  at sufficiently large $V$.

\section{Models and methods}
\subsection{Hubbard model}
To begin, we briefly review the MIT on bipartite lattices in the Hubbard model [Eq. (\ref{eq:effHub})], i.e., without nonlocal interactions. For the systems here, in $d>2$ the local interaction $U$ induces a MIT from a paramagnetic metal to an antiferromagnetic insulator \cite{gebhard_florian_mott_1997}. For lattices with perfect nesting the critical interaction $U_c$ vanishes for zero temperatures \cite{georges_dynamical_1996}. For lattices with a vanishing density of states (DOS) at the Fermi energy $E_F$, $U_c$ is finite for $T=0$ \cite{santoro_hubbard_1993}. In two dimensions the Mermin-Wagner theorem prevents long-range order and therefore a magnetic transition \cite{walker_absence_1968,ghosh_nonexistence_1972,koma_decay_1992}. However, quasi-long-range antiferromagnetic fluctuations lead to a similar phase diagram for the transition from a paramagnetic metal to a quasi-ordered insulator. Again, perfect nesting in the square lattice leads to $U_c=0$ for $T=0$ \cite{schafer_fate_2015}, while the vanishing DOS at $E_F$ in the honeycomb lattice leads to $U_c>0$ for $T=0$ \cite{sorella_absence_2012,assaad_pinning_2013}. In all cases the gap appears continuously; that is, the MIT is not of first order.
\subsection{Extended Hubbard model}

We now turn to the extended Hubbard model,
\begin{eqnarray}
H = -t\sum_{\langle i,j\rangle,\sigma} c_{i\sigma}^\dagger c^{\phantom{\dagger}}_{j\sigma} + U\sum_i n_{i\msu}n_{i\msd} + \frac{1}{2} \sum_{i\neq j}V_{ij} n_in_j, \label{eq:exHub}
\end{eqnarray}
where $c_{i\sigma}^{(\dagger)}$ is the annihilation (creation) operator for electrons on site $i$ and spin $\sigma$, $t$ is the nearest-neighbor hopping amplitude, $U$ is the local interaction, and $V_{ij}$ is the nonlocal interaction between electrons at sites $i$ and $j$. $n_{i\sigma}$ and $n_i$ are the spin-resolved and total occupation operators, respectively. We focus on nearest-neighbor (NN), $V_{0j}=V \delta_{01}$, and long-range (L), $V_{0j}=V/r_j$, interactions \footnote{The $r_j$ are scaled such that $V_{01}=V$.}.

In this model repulsive nonlocal interactions decrease correlations \cite{schuler_optimal_2013,hohenadler_phase_2014,wu_phase_2014,2016slft.confE.244B,van_loon_capturing_2016,terletska_charge_2017,ayral_influence_2017} and can lead into a CDW phase \cite{hirsch_charge-density-wave_1984}. Our focus is solely on how nonlocal interactions influence the order of the MIT.

\subsection{Variational principle}
\begin{figure}[h]
\mbox{
\includegraphics[width=0.99\columnwidth]{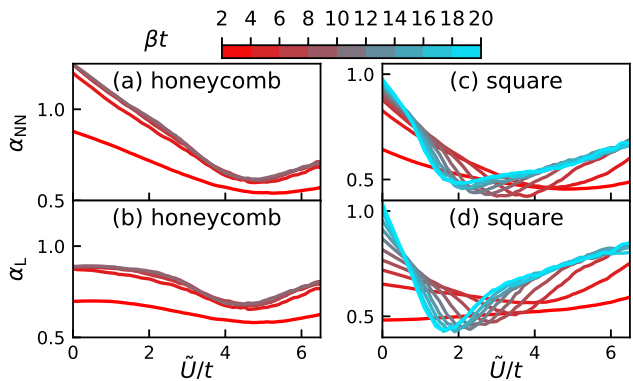}
}
\caption{(Color online) Effective screening factor $\alpha$ as defined above Eq. (\ref{eq:ustar_short}) for NN (top panels) and L (bottom panels) interaction. Left (right) panels show the case of the honeycomb (square) lattice. We show results from low (red) to high (cyan) inverse temperatures $\beta t$.}
\label{fig:alpha}
\end{figure}

We investigate the $U$-$V$-$T$ phase diagram of the extended Hubbard model by approximating its thermodynamic ground state using the Peierls-Feynman-Bogoliubov variational principle~\cite{Peierls38,Bogoliubov58,Feynman72} with a Hubbard model as the effective system~\cite{schuler_optimal_2013}. The effective Hubbard model, reading
\begin{eqnarray}
\tilde H = -t\sum_{\langle i,j\rangle,\sigma} c_{i\sigma}^\dagger c_{j\sigma}^{\phantom{\dagger}} + \tilde U\sum_i n_{i\msu}n_{i\msd},  \label{eq:effHub}
\end{eqnarray}
is varied via the effective local interaction $\tilde U$ in order to minimize a free energy functional. Therefore $\tilde U$ is
\begin{eqnarray}
\tilde U =U+\sum_{j\neq 0}V_{0j}  \frac{\partial_{\tilde U}\langle{n_{0}n_{j}}\rangle_{\tilde H}}{\partial_{\tilde U} \langle n_{0}n_{0}\rangle_{\tilde H}}. \label{eq:ustar}
\end{eqnarray}
Although the variational principle provides only an upper bound of the free energy, it has been found to give an accurate description of the physics present in the effective reference Hubbard model \cite{van_loon_capturing_2016} and even
gives exact double occupancies for infinitesimal nonlocal interactions \cite{van_loon_capturing_2016}. 
This makes the approach appropriate for capturing the MIT, a hallmark of Hubbard model physics, in the extended Hubbard model. We introduce the effective screening factors
$\alpha_\text{NN}(\tilde U)=-\frac{\partial_{\tilde U}\langle{n_{0}n_{1}}\rangle_{\tilde H}}{\partial_{\tilde U} \langle n_{0}n_{0}\rangle_{\tilde H}} $
and $ \alpha_\text{L}(\tilde U)=-\sum_{j\neq 0}\frac{1}{r_j} \frac{\partial_{\tilde U}\langle{n_{0}n_{j}}\rangle_{\tilde H}}{\partial_{\tilde U} \langle n_{0}n_{0}\rangle_{\tilde H}} $
with which Eq. (\ref{eq:ustar}) simplifies to
\begin{eqnarray}
\tilde U = U-V \alpha_\text{NN/L}(\tilde U), \label{eq:ustar_short}
\end{eqnarray}
for the case of NN and L interactions. Here, $\alpha(\tilde U)$ is a property of the effective Hubbard model and quantifies the above mentioned decrease of correlations by $V$: Nonlocal interactions shift the transition at $\tilde U_\text{MIT}$ to leading order linearly with a slope of $\alpha^{-1}(\tilde U =\tilde  U_\text{MIT})$; that is, a positive $\alpha$ leads to nonlocal interactions stabilizing the metallic phase. Indeed, we find strictly positive $\alpha$ in our numerical calculations (see Fig.~\ref{fig:alpha}). 

From Eq. (\ref{eq:ustar_short}) we calculate the change in the effective interaction $\tilde U$ with $V$ and $U$, 
\begin{eqnarray}
\left(
\partial/\partial U, \, \partial/\partial V
  \right)\tilde U =  \left(1+V\frac{\partial \alpha}{\partial \tilde U} \right)^{-1} \left( 1, \,  -\alpha
\right). \label{eq:divergence}
\end{eqnarray}
The derivatives diverge for $-1=V_c \partial_{\tilde U} \alpha $. At this point we find a bifurcation of the solution of Eq. (\ref{eq:ustar_short}); that is, a single point $(U,V)$ is mapped to multiple $\tilde U$ and thus a jump of the observables of the effective system. Particularly, discontinuities in $\alpha(\tilde U)$ lead to arbitrarily small nonlocal Coulomb interactions inducing a first-order phase transition ($V_c=0$). 

In the following we numerically determine the critical $V_c$ required to induce such a first-order phase transition. To this end we calculate $\alpha$ for different lattice topologies and dimensions: We choose to investigate the square and honeycomb lattices and their three-dimensional generalizations the cubic and diamond lattices. We rely on quantum Monte Carlo simulations of all effective Hubbard models.

For the case of two dimensions we use the determinant quantum Monte Carlo method (DQMC) \cite{blankenbecler_monte_1981} implemented in the \textsc{quest} code \footnote{``QUantum Electron Simulation Toolbox'' \textsc{quest} 1.3.0 A. Tomas, C-C. Chang, Z-J. Bai, and R. Scalettar, (\url{http://quest.ucdavis.edu/})}. We alleviate finite-size and Trotter errors by extrapolating from finite Trotter discretizations of $\Delta \tau =0.2$, $0.1$, and $0.05$ and linear lattice sizes of $L=8$, $10$, and $12$ for the square lattice and $L=6$, $9$, and $12$ for the honeycomb lattice \cite{rost_momentum-dependent_2012}. We evaluate derivatives in the definition of $\alpha$ numerically by solving Hubbard models in steps of $\Delta \tilde  U/t = 0.1$ and deal with statistical noise by a Savitzky-Golay approach. We provide raw data and details on the DQMC simulations, finite-size and Trotter extrapolations, and the Savitzky-Golay approach in Appendix  \ref{sec:app_DQMC}.
\section{Results}
\subsection{Honeycomb lattice}
First, we discuss the case of the honeycomb lattice for which the $\alpha(\tilde U)$ are plotted for $\beta t = 2$ to $\beta t = 10$ in Figs.~\ref{fig:alpha}(a) and (b). For $\beta t \gtrsim 6 $ no temperature dependence is observable due to the linearly vanishing DOS at $E_F$ \cite{wu_interacting_2010}. Thus, we can draw conclusions for finite temperatures \textit{and} $T\rightarrow 0$. For low temperatures $\alpha$ shows a minimum at $\tilde U_\text{min} /t \sim 4.6$ and a maximal absolute derivative at $\tilde U_\text{kink}/t \sim 3.2$. The MIT is in between at $\tilde U_\text{MIT}/t\approx 3.8$ \cite{sorella_absence_2012,assaad_pinning_2013}.

\begin{figure}[b]
\mbox{
\includegraphics[width=0.99\columnwidth]{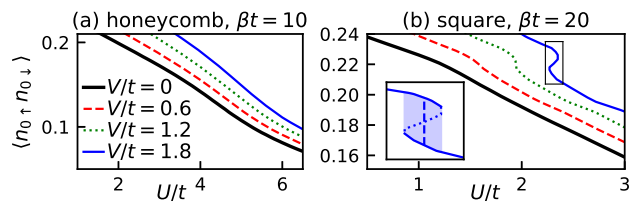}
}
\caption{(Color online) Double occupancy for (a) the honeycomb ($\beta t = 10$) and (b) square lattice ($\beta t = 20$) for $V/t=0$ (black), $0.6$ (dashed red), $1.2$ (dotted green), and $1.8$ (thin blue) for the case of NN interaction. Inset in (b) shows a close up of the rectangle with the thermodynamically unstable states (dotted), coexistence region (shaded), and double occupancy by Maxwell construction (dashed).}
\label{fig:dd}
\end{figure}

Notably, the dependence of the effective screening factor $\alpha$ on $\tilde U$ is rather weak (no steplike features). Hence, there is no $\tilde U$ where the slope $\partial \alpha/\partial \tilde U$ is particularly large, and from Eq. (\ref{eq:divergence}) we expect that rather large $V_c$ would be needed to push the MIT to first order here. Clarifying this, we solve Eq. (\ref{eq:ustar_short}) and calculate the $U$-dependent double occupancy of an extended Hubbard model with different nearest-neighbor interaction at $\beta t =10$ [Fig.~\ref{fig:dd}(a)]. For increasing $V$ the $V=0$ line is shifted towards larger $U$ (i.e., $V$ weakens correlations).

Concerning the influence of nonlocal interactions on the \textit{order} of the transition, we calculate $V_c$ from the slope of $\alpha$. We find $V_c/t\approx 7.7$ and $V_c/t\approx 14.3$ for NN  and L, respectively. This is in line with findings that the transition is continuous up to $V/t\sim 1.5$ in the case of nearest-neighbor interaction \cite{wu_phase_2014} since larger $V$ stabilize a CDW phase which we estimate in strong coupling, as presented in Appendix \ref{sec:app_strong_coupl}. For the honeycomb lattice $V_c$ is always larger than $V_\text{CDW}$ such that no first-order MIT will be observable.

We infer a $U$-$V$ phase-diagram schematically shown in Fig.~\ref{fig:schem}(b). The slope of the transition line at $V=0$ is given by $1/\alpha(U_\text{MIT})$, with $\alpha_\text{NN(L)}(U_\text{MIT}) = 0.66 (0.60)$. Quantum Monte Carlo (QMC) calculations with long-range interactions \cite{hohenadler_phase_2014} reveal a slope compatible to $\alpha_\text{L} \lesssim 0.55$; For nearest-neighbor interactions dynamical cluster approximation (DCA) calculations \cite{wu_phase_2014} indicate $\alpha_\text{NN} \sim 0.23$, whereas QMC calculations \cite{2016slft.confE.244B} lead to $\alpha_\text{NN} \lesssim 0.52$.

\subsection{Square lattice}
The case of the square lattice turns out to be different. We show $\alpha_\text{NN/L}$ in Figs.~\ref{fig:alpha}(c) and (d) for $\beta t = 2$ to $\beta t = 20$. Here, $\alpha$ is strongly temperature dependent. Prominently, $\tilde U_\text{min}$ and $\tilde U_\text{kink}$ are shifted to smaller $\tilde U$, and the slope at $\tilde U_\text{kink}$ gets steeper with increasing $\beta$. The increase in the slope of $\alpha(\tilde U)$ traces back to the development of a soft kink in $\langle n_0 n_0 \rangle(\tilde U) $. Comparing $U_\text{min}$ and $U_\text{kink}$ to the temperature dependent critical interaction $\tilde U_\text{MIT}$ from Ref.~\onlinecite{schafer_fate_2015}, we find that $\tilde U_\text{min} \approx\tilde  U_\text{MIT}$ and that $\tilde U_\text{kink}$ approaches $\tilde U_\text{MIT}$ with lower temperatures.

Concerning the resulting phase diagram [Fig.~\ref{fig:schem}(a)], the slope of the $U$-$V$ phase-transition line at $V=0$  is $1/\alpha(\tilde U_\text{MIT})$, with $\alpha_\text{NN(L)}( \tilde U_\text{MIT}) \sim 0.5$ for all temperatures. Results for nearest-neighbor interaction from DCA \cite{terletska_charge_2017} at $\beta t = 6$ and combined GW plus extended DMFT \cite{ayral_influence_2017} at $\beta t = 25$ are compatible with $\alpha\sim 0.8$ and $\alpha \sim 0.62$.

\begin{figure}[h]
\mbox{
\includegraphics[width=0.99\columnwidth]{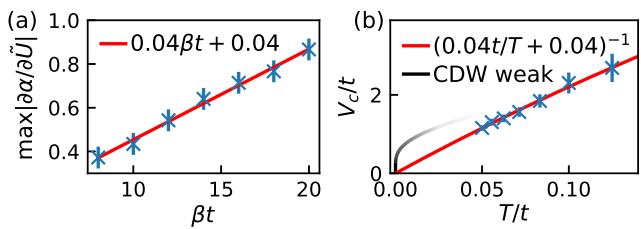}
}
\caption{(Color online) (a) Dependence of  $\max[|\partial \alpha / \partial \tilde U |]$ on $\beta t$ for the square lattice with linear fit (red line). (b) Corresponding $V_c(T)=-[\partial \alpha / \partial \tilde U]^{-1}$. Results for $\mathrm{NN}$ and $\mathrm{L}$ coincide within error bars. The fading black curve is a weak-coupling estimate of the critical $V$ for the CDW valid for small $V$.}
\label{fig:alphaDerivMin}
\end{figure}

In order to estimate the smallest nonlocal interaction $V_c$ leading to a first-order transition we calculate $\max |\partial_{\tilde U}\alpha (\tilde U)|$, which turns out to exhibit a linear $\beta$ dependence, as can be seen in Fig.~\ref{fig:alphaDerivMin}(a). Thus, by extrapolating to decreasing temperatures [Fig.~\ref{fig:alphaDerivMin}(b)] we expect a first-order phase transition at decreasing nonlocal interaction strengths: $V_c \rightarrow 0$ for $T\rightarrow 0$. Here, we find $V_c/t \approx 1.15$ for $\beta t = 20$. For NN interaction the first-order phase transition is probably not observable for $\beta t=20$ since it is deeply buried in the CDW phase (see Appendix \ref{sec:app_strong_coupl}). However, since $V_c$ scales linearly with $T$ but the nonlocal interaction for the CDW phase scales as $V_\text{CDW}/t=4 \pi^2 \left[\ln (8{t}/{T})\right]^{-2} $ \cite{fazekas_lecture_1999}, [black line in Fig.~\ref{fig:alphaDerivMin}(b)], low enough temperatures \textit{always} lead to a favoring of the first-order MIT over the CDW. Moreover, long-range interactions partially suppress the CDW phase such that in this case the first-order MIT will be observable at only slightly lower temperatures (see Appendix \ref{sec:app_strong_coupl}).

In Fig.~\ref{fig:dd}(b) we present the double occupancy dependent on $U$ at $\beta t =20$ for different nonlocal interactions. The curves are shifted to larger $U$ with increasing $V$, where the different amounts of shifting are apparent for the metallic and insulating regimes. This different effective screening on the metallic and insulating sides of the transition eventually lifts the MIT to first order here. The soft kink visible for $V=0$ (black solid line) gets a steplike shape for $V>0$, which for $V > V_c \approx 1.2 t$ eventually leads to unphysical (see below) loops, as shown in detail in Fig.~\ref{fig:dd}(d) for $V=1.8t$. The real double occupancy in the coexistence region is obtained by Maxwell construction and shown as a dashed line. This coexistence region is shown schematically in Fig.~\ref{fig:schem}(a).

The double occupancy $D=\langle n_{0\msu} n_{0\msd}\rangle$ and the local Hubbard interaction $U$ are conjugate variables in the thermodynamic sense, i.e., $D=\frac{1}{N}\partial F/\partial U$, where $F$ is the free energy and $N$ is the number of lattice sites in the system. Since $F$ is not only extremal but actually minimal in a stable thermodynamic state, a small deviation from the thermodynamic ground state must increase the free energy. The resulting thermodynamic inequality $\partial D/\partial U<0$ demands that the double occupancy is a monotonously decreasing function of the on-site interaction, as detailed in Appendix \ref{sec:app_thermo}.

This inequality is fulfilled everywhere except for the dashed part of the $D(U)$ curve inside the hysteresis region [Fig.~\ref{fig:dd}(d)], which thus corresponds to thermodynamically unstable states. This behavior is characteristic of a first-order transition and signals that the metallic and the insulating sides of the transition are not linked continuously through a series of thermodynamically stable states.

\subsection{Cubic and diamond lattices}
We now turn to higher-dimensional systems with cubic \cite{jarrell_hubbard_1992} and diamond \cite{santoro_hubbard_1993} lattices, which generalize the square and honeycomb lattice to three dimensions. While the diamond lattice preserves the linearly vanishing DOS at $E_F$, the cubic lattice loses the van Hove singularity at $E_F$ but keeps a nonzero DOS at $E_F$. 
The main difference with $d=2$ is the absence of the Mermin-Wagner theorem and the presence of finite-temperature antiferromagnetic long-range order. We solve the Hubbard models in $d=3$ in DMFT  using \textsc{triqs} \cite{parcollet_triqs:_2015,seth_triqs/cthyb:_2016}. We allow for antiferromagnetic long-range order to study the thermodynamically relevant transition from a (semi)metal to an antiferromagnetic insulator \cite{georges_dynamical_1996}. We provide raw data, details on the simulations, and results for the case of infinite dimensions in Appendix \ref{sec:app_DMFT}. From DQMC simulations in $d=3$ we find that the discontinuous behavior of $\alpha$ is essentially determined by $\partial_{\tilde U}\langle n_{0\msu} n_{0\msd} \rangle$ (see Appendix \ref{sec:app_DQMC_cubic} for details) and thus search for discontinuities in $\partial_{\tilde U}\langle n_0 n_0 \rangle $ directly, circumventing the calculation of $\partial_{\tilde U}\langle n_0 n_j \rangle $ for $j>0$.

\begin{figure}[h]
\mbox{
\includegraphics[width=0.99\columnwidth]{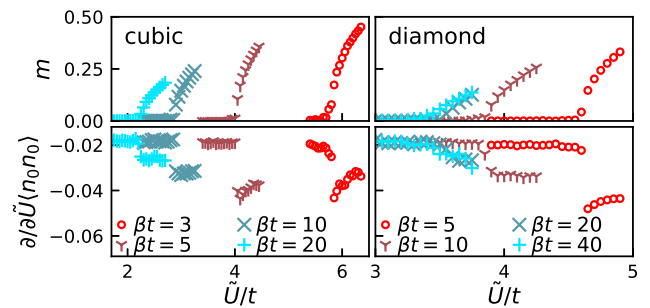}
}
\caption{(Color online) DMFT results for the cubic (left panels) and diamond (right panels) lattices. From top to bottom the panels show staggered magnetization $m$ and $\partial_{\tilde U } \langle n_0 n_0 \rangle$.}
\label{fig:d_3}
\end{figure}

The results for different temperatures  are presented in Fig.~\ref{fig:d_3}. The onset of a finite staggered magnetization $m$ determines the critical effective interaction $\tilde U_c$ at the MIT. For the diamond lattice, $\tilde U_c$ becomes temperature independent for small enough temperatures \cite{santoro_hubbard_1993} like in $d=2$. For the cubic lattice $U_c$ approaches zero for $T \rightarrow 0$ \cite{rohringer_critical_2011}. We find clear discontinuities in the double occupancies' derivative and thus find discontinuities in $ \alpha$ for all temperatures only for the cubic lattice. The size of the discontinuity grows as the system approaches large N\'eel temperatures \cite{jarrell_hubbard_1992,santoro_hubbard_1993}. From this discontinuity we conclude that infinitesimal positive nonlocal interactions induce a first-order phase transition in three or more dimensions; that is, we expect $V_c=0$ for $d>2$ in cubic systems. The linearly vanishing DOS at $E_F$ in the case of the diamond lattice leads to no discontinuities at low temperatures. For large enough temperatures ($\beta t \lesssim 10$) the linearly vanishing DOS is smeared out, such that a discontinuity in $\partial_{\tilde U}\langle n_0 n_0 \rangle$ appears. We conclude that for low temperatures only finite nonlocal interactions induce first-order phase transitions in diamondlike lattices in arbitrary dimensions.

This dimensional dependence of the MIT in the cubic systems can be understood from the nature of the antiferromagnetism. An antiferromagnetic phase transition translates to a kink in the double occupancy since the latter is related to the magnetic moment as $m^2 =n - 2n_\msu n_\msd$. In $d=2$, the Mermin-Wagner theorem forbids conventional (i.e., second order in the Ehrenfest sense) antiferromagnetic phase transitions at finite temperature, which leads to a smooth double occupancy and a finite $V_c$. For $d>2$, on the other hand, an antiferromagnetic phase transition \cite{gebhard_florian_mott_1997, rohringer_critical_2011} leads to a kink in the double occupancy and a first-order phase transition at infinitesimal $V$. The vanishing DOS for the diamond lattice, on the other hand, leads to an unusual critical behavior \cite{sorella_semi-metal-insulator_1992}, and thus no kink in the double occupancy and a finite $V_c$.

\section{Conclusion}
We showed that nonlocal interactions in bipartite extended Hubbard models can lead to a first-order MIT. This result is highly relevant in the context of the question of whether Hubbard models describe discontinuous MITs occurring in realistic materials. The underlying mechanism is governed by nonlocal interactions screening correlations differently in the insulating and metallic phases, with the metallic phase being generally stabilized by the nonlocal interactions. Interestingly, this is in contrast to the mechanism envisioned by Mott \cite{mott_basis_1949}, which is based on nonlocal interactions stabilizing the \textit{insulating} phase. We found first-order transitions for nonlocal interactions larger than a critical $V_c(U)$. Our calculations indicate $V_c=0$ for cubic systems in $d>2$, whereas systems with vanishing DOS at $E_F$ (e.g., diamond) and two-dimensional systems in general show $V_c>0$ for low temperatures. With nonlocal interactions, we found an additional mechanism, next to lattice distortions and multi-orbital physics \cite{ono_mott_2003,pruschke_hunds_2005,facio_nature_2017}, explaining how the continuous MIT in Hubbard models is reconvened with the discontinuous MIT in real materials.

\textit{Acknowledgments.} 
We acknowledge the kind help of R. Staudt. We thank S. Haas, S. Wessel, A. Rosch, and F. Gebhard for comments. E.G.C.P.v.L. and M.I.K. acknowledge support from ERC Advanced Grant No. 338957 FEMTO/NANO. Computer time at the HLRN is acknowledged.

\appendix

\section{Details of DQMC simulations}
\label{sec:app_DQMC}
We perform simulations at a fixed value of $\tilde{U}$ and at half filling, i.e., by setting $\mu=\tilde{U}/2$. As discussed in Ref.~\onlinecite{scalettar_ergodicity_1991}, simulations of the Hubbard model for $U /t \gtrsim 8$ require global updates in order to explore the phase space in an ergodic manner. Although we restrict our simulations to $U/t\lesssim 6$, we include global moves as a precautionary measure and indeed find no ``sticking'' behavior of the occupancies described in Ref.~\onlinecite{scalettar_ergodicity_1991}. With  500 warm-up sweeps we perform between 10000 and 1 million measurement sweeps depending on the temperature, lattice size, and Trotter discretization.  We provide our raw data together with more equal time measurements provided by the \textsc{quest} software for all lattice sizes, Trotter discretizations, and temperatures for both the square and honeycomb lattices together with error estimates and a complete set of input parameters on the Zenodo platform \cite{schuler_determinant_2017}.

\begin{figure}[h]
	\mbox{
		\includegraphics[width=0.99\columnwidth]{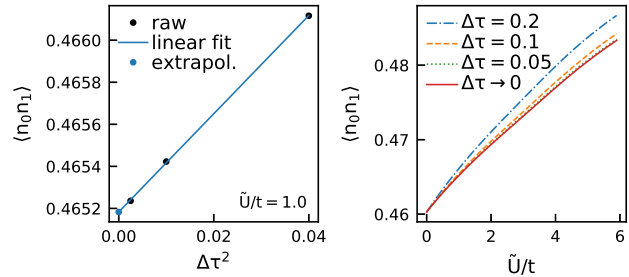}
	}
	\caption{(Color online) Finite-$\Delta \tau$ extrapolation for the nearest neighbor charge correlator on the square lattice for $\beta t = 10$ and $L=10$. Left: linear fit (blue line) to raw data on $\Delta \tau ^{-2}$ (black dots) resulting in extrapolated value at $\Delta \tau \rightarrow 0$ (blue dot) for $\tilde U / t =1.0$. Right: $\tilde U$-dependent results for the extrapolation.}
	\label{fig:finite_dtau}
\end{figure}

\begin{figure}[h]
	\mbox{
		\includegraphics[width=0.99\columnwidth]{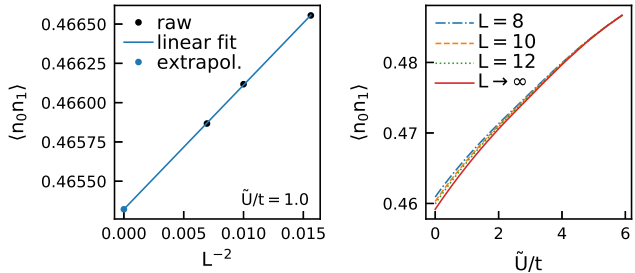}
	}
	\caption{(Color online) Finite-$L$ extrapolation for the nearest neighbor charge correlator on the square lattice for $\beta t = 10$ and $\Delta \tau=0.2$. Left: linear fit (blue line) to raw data on $L^{-2}$ (black dots) resulting in the extrapolated value at $\Delta \tau \rightarrow 0$ (blue dot) for $\tilde U / t =1.0$. Right panel: $\tilde U$-dependent results for the extrapolation.}
	\label{fig:finite_L}
\end{figure}

We deal with finite-size and finite Trotter errors by extrapolating schemes: We simulate effective Hubbard models in $d=2$ with  imaginary time discretizations $\Delta \tau =0.05$, $ 0.1$, and $0.2$ and extrapolate a linear dependence on $\Delta \tau ^2$. We present an example of this procedure in Fig.~\ref{fig:finite_dtau}. The errors for finite $\Delta \tau $ do not exist for $\tilde U = 0$ and increase for larger $\tilde U$. From linear system sizes $L=8$, $10$, and $12$ for the square lattice and $L=6$, $9$, and $12$ for the honeycomb lattice we extrapolate a linear dependence on $L^{-2}$. We show an example in Fig.~\ref{fig:finite_L}. 
The finite-size errors in the charge correlation function get smaller for larger $\tilde U$ since the electrons localize.

\begin{figure}[h]
	\mbox{
		\includegraphics[width=0.99\columnwidth]{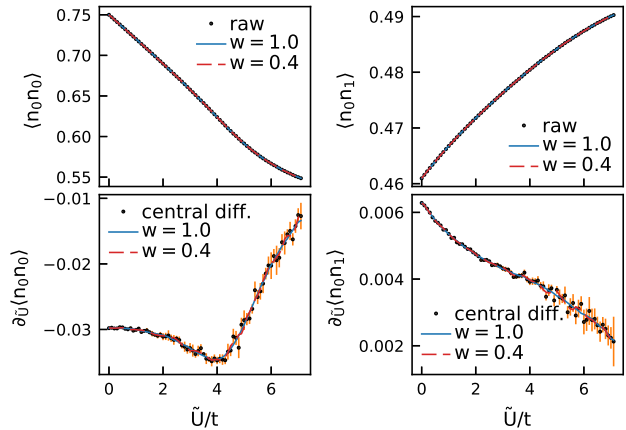}
	}
	\caption{(Color online) Raw data and smoothed data for on-site and nearest-neighbor charge correlators ($\langle n_0 n_0\rangle$, $\langle n_0 n_1\rangle$; top panels) and their derivatives with respect to $\tilde U$ (bottom panels) for the square lattice with $L=12$, $\Delta \tau = 0.2$, and $\beta t= 4.0$. We show smoothing results of different fit windows, $w=1.0$ and $w=0.4$, with cubic polynomials. Where no error bars are visible, they are hidden behind the markers. }
	\label{fig:sav_1}
\end{figure}

\begin{figure}[h]
	\mbox{
		\includegraphics[width=0.99\columnwidth]{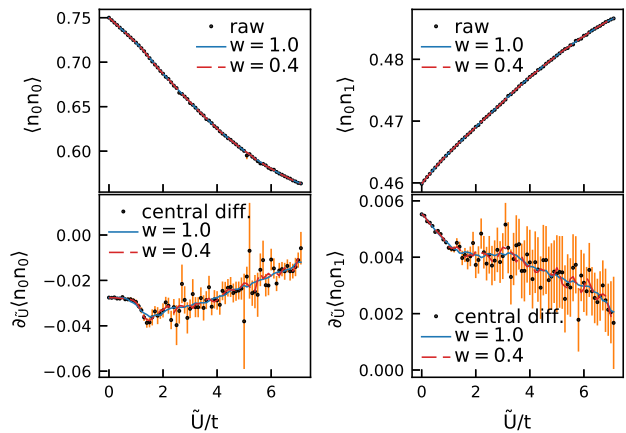}
	}
	\caption{(Color online) Same as Fig.~\ref{fig:sav_1}, but for $L=12$, $\Delta \tau = 0.05$, and $\beta t= 20.0$.}
	\label{fig:sav_2}
\end{figure}

In order to reduce the inherent noise in the Monte Carlo data, which poses a serious problem when calculating derivatives with respect to $\tilde{U}$, we
use a Savitzky-Golay approach \cite{savitzky_smoothing_1964}; that is, we analytically take derivatives of polynomials which are locally fitted in a window of width $w$ to the numerical values of $\langle n_{0} n_{j}\rangle(\tilde U)$. We show an example of this procedure for two different cases: smooth dependence on $\tilde U$ with little noise for high temperatures ($\beta t=4.0$; Fig.~\ref{fig:sav_1}) and rather large dependence on $\tilde U$ with large noise for low temperatures ($\beta t=20.0$; Fig.~\ref{fig:sav_2}). We show results for different fit windows ($w=0.4$ and $w=1.0$) for cubic polynomials. In all cases, the raw data and the smoothed data are hard to distinguish. However, the derivative with respect to $\tilde U$ vastly increases noise for the raw data. Taking the derivative analytically for the smoothed data avoids this. The high-temperature data show that the larger window leads to smoother results. The case of low temperature, however, exemplifies the drawback of too large windows: The steep feature in $\partial_{\tilde U}\langle n_0 n_0 \rangle$ at $\tilde U /t \sim 1.3$, which can be clearly seen in the raw data, is washed out for $w=1.0$. The smaller window $w=0.4$ leads to data which nicely follow the raw data for small to intermediate $\tilde U /t $ but also shows larger noise for larger $\tilde U /t$. Since the derivative of $\partial_{\tilde U}\langle n_0 n_0 \rangle$ (via that of $\alpha$) determines the critical $V_c$, we have extracted them with $w=0.4$ and cubic polynomials, which is shown in Fig.~\ref{fig:alphaDerivMin}. The effective screening factors shown in Fig.~\ref{fig:alpha} are obtained with a window of $\tilde U_w/t=1.0$ and quadratic polynomials, such that the strong noise at large $\tilde U /t$ does not obstruct the trends visible in Fig.~\ref{fig:alpha}.

A word on the error bars shown in Figs.~\ref{fig:sav_1} and \ref{fig:sav_2}: The quantities (with error bars) actually measured in the DQMC algorithm are $\langle n_{0\msu} n_{i\msd}\rangle$ and $\langle n_{0\msu} n_{i\msu}\rangle$, such that we obtain $\langle n_0 n_i \rangle$ by summing over the two observables. Since the two constituent observables are correlated (for $i\neq 0$), the error bar on their sum cannot simply be obtained by Pythagorean addition.
Visual inspection of the raw data \cite{schuler_determinant_2017} at neighboring values of $\tilde{U}$ shows that error cancellation happens in the determination of $\langle n_0 n_i \rangle$ (equivalently, the statistical errors in $\langle S^z_0 S^z_i \rangle$ are larger than the Pythagorean sum of the error bars of $\langle n_{0\msu} n_{i\msd}\rangle$ and $\langle n_{0\msu} n_{i\msu}\rangle$  ).
In the figures, we have done Pythagorean addition and scaled the errors, such that the error bars enclose 70\% of the smoothed data, which leads to visually reasonable results but overall too large error bars for $\tilde U/t \gtrsim 3$. Since the error estimates do not influence the calculations, this is, however, not a crucial point.

\section{Details of DMFT simulations}
\label{sec:app_DMFT}
For our calculations we use the \textsc{triqs} package \cite{parcollet_triqs:_2015} and the accompanying continuous-time quantum Monte Carlo hybridization expansion solver \cite{seth_triqs/cthyb:_2016}.
We find a critical slow down of the DMFT convergence on the paramagnetic side of the antiferromagnetic transition. For some cases we use more than 400 DMFT cycles to obtain reasonable convergence. We subsequently perform some iterations with increased statistics with up to 4 million sweeps on 80 cores each to obtain data with little noise. We cannot rely on a smoothing algorithm as in the case of finite-size DQMC data since the kink in the double occupancy would vanish with any smoothing algorithm. We calculate the derivative $\partial_{\tilde U} \langle n_0 n_0 \rangle$ by performing forward and backward finite differences. If both (forward and backward differences) are equal within a tolerance of 2\%, we take the mean value (i.e., we perform central difference); if they are not, we assume the derivative is not defined at that $U$ [i.e., there is a kink in $\langle n_0n_0 \rangle(\tilde U)$]. 

\begin{figure}[h]
	\mbox{
		\includegraphics[width=0.99\columnwidth]{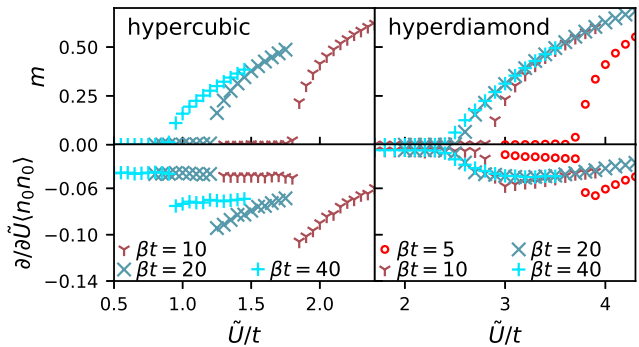}
	}
	\caption{(Color online) DMFT results for the $d= \infty$ hypercubic (left panels) and hyperdiamond (right panels) lattices. From top to bottom the panels show staggered magnetization $m$ and $\partial_{\tilde U } \langle n_0 n_0 \rangle$.}
	\label{fig:d_infty}
\end{figure}

We present DMFT results similar to the DMFT simulations in $d=3$ presented in Fig.~\ref{fig:d_3} for $U$ values close to the antiferromagnetic phase transition for the corresponding lattices in $d=\infty$, i.e., the hypercubic \cite{jarrell_hubbard_1992} and hyperdiamond \cite{santoro_hubbard_1993} lattices. The case of $d=\infty$ is interesting since DMFT provides an exact solution. The results are presented in Fig.~\ref{fig:d_infty} in the same way as for $d=3$ above. We have performed calculations for $\beta t=10$, $20$, and $40$ and $\beta t=5$, $10$, $20$, and $40$ for the hypercubic and hyperdiamond lattices, respectively. The results are qualitatively very similar to the case of $d=3$; that is, we find discontinuities in $\partial_{\tilde U} \langle n_0 n_0 \rangle $ for all temperatures for the hypercubic case. For the hyperdiamond case we find a discontinuity only at the highest temperature ($\beta t = 5$). The different values for the critical $U$ in $d=3$ and $d=\infty$ can be understood in terms of different bandwidths. Using effective half bandwidths of $2.2t$ and $2.8t $ for the hypercubic and hyperdiamond lattice, respectively, the values of $U_c/w$ ($w$ is the bandwidth) for $d=3$ and $d=\infty$ match nearly perfectly. 

We provide raw data (Greens function and self-energy of Matsubara frequencies, occupancy, and double occupancy)  for the last DMFT iteration for all temperatures and all four lattices [ cubic, diamond, hypercubic ($d=\infty$), and hyperdiamond ($d=\infty$)] together with a complete set of input parameters on the Zenodo platform \cite{schuler_dynamical_2017}.

\section{DQMC results for the cubic lattice and comparison with DMFT}
\label{sec:app_DQMC_cubic}
\begin{figure}[h]
	\mbox{
		\includegraphics[width=0.99\columnwidth]{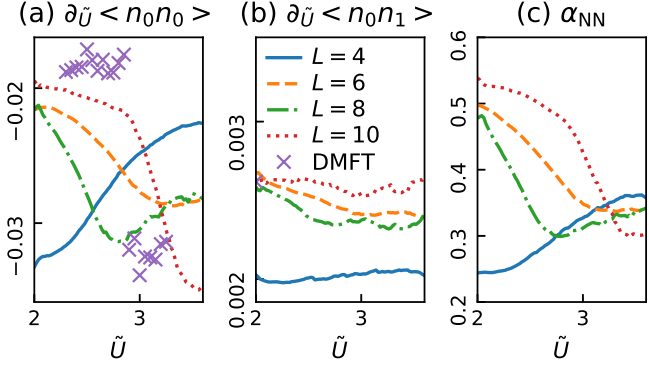}
	}
	\caption{(Color online) Results for the derivative of the (a) local and (b) nearest-neighbor charge correlator with respect to $\tilde U$ and (c) $\alpha_\text{NN}$ as defined in the main text from DQMC simulation of the cubic lattice at $\beta t =10$ for lattices with linear sizes $L=4$ (solid blue line), $L=6$ (dashed yellow line), $L=8$ (dash-dotted green line), and $L=10$ (dotted red line). Results for the local correlator in the thermodynamic limit $L\rightarrow\infty$ in the DMFT approximation are presented as purple crosses in (a). }
	\label{fig:dqmc_dmft}
\end{figure}

A DQMC treatment of the Hubbard model on the cubic lattice suffers from the scaling of computational time with the linear lattice size, which is $\propto L^9$ and limits the calculations to $L\le 10$. To assess the finite-size scaling we have performed simulations for $\beta t= 10$ at fixed Trotter discretization of $\Delta \tau=0.1$. We do not perform an extrapolation to $\Delta \tau=0$ since our results for the square lattice and test calculations at $\beta t = 4$ for the cubic lattice show that results for $\Delta \tau=0.1$ are reasonably close to the extrapolated value (see Fig.~\ref{fig:finite_dtau}). In Fig.~\ref{fig:dqmc_dmft} we present calculations for $L=4$, $6$, $8$, and $10$ for interaction strengths between $\tilde U/t=2$ and $\tilde U/t=3.78$ in steps of $0.02$ in terms of derivatives of the local and nearest-neighbor charge correlators with respect to $\tilde U$ as well as $\alpha_\text{NN}$. We smooth the data with $w=0.3$ and quadratic polynomials. From the analysis of the data we will answer two questions: First, does a discontinuity in $\partial_{\tilde U} \langle n_0 n_0\rangle$ translate into a discontinuity in $\alpha_\text{NN}$, or is it canceled by a respective discontinuity in $\partial_{\tilde U} \langle n_0 n_1\rangle$? Second, what do we learn from the comparison of DMFT and DQMC data?

To answer the first question, we investigate the finite-size scaling of $\partial_{\tilde U} \langle n_0 n_0\rangle$ and $\partial_{\tilde U} \langle n_0 n_1\rangle$. As can be seen from, e.g., the position of the minimum of $\partial_{\tilde U} \langle n_0 n_0\rangle$, the finite-size behavior is nonmonotonous, which makes an extrapolation to $L\rightarrow \infty$ impossible based on these data. However, we can observe that the step-like feature visible in $\partial_{\tilde U} \langle n_0 n_0\rangle$ gets monotonously sharper for larger $L$. For the nearest-neighbor case, no steplike feature is observed for any $L$. Consequently, we find steplike features in $\alpha_\text{NN}$ which get sharper in the same way as they do for the derivative of the local correlator. If we take for granted that this behavior holds for $L=\infty$, i.e., discontinuities exist only for the on-site case and not for the nearest-neighbor case, a discontinuity in $\partial_{\tilde U} \langle n_0 n_0\rangle$ directly translates to one in $\alpha_\text{NN}$ and thus signals a first-order phase transition for arbitrarily small nonlocal interactions.

The result of the DMFT solution of the cubic lattice is presented in Fig.~\ref{fig:dqmc_dmft} (a). Although DMFT provides only an approximate solution for the cubic lattice, it does give a result in the thermodynamic limit and can thus be seen as a crude finite-size extrapolation of the DQMC data. From a superficial inspection of the data the finite-size DQMC data seem to converge against the DMFT result. A detailed comparison of DMFT with DCA\cite{fuchs_thermodynamics_2011}, dynamical vertex approximation (D$\Gamma$A) \cite{rohringer_critical_2011}, and DQMC\cite{staudt_phase_2000} for the cubic Hubbard model suggest that for small interaction strengths ($U/t \lesssim 3.5 $) the DMFT result coincides rather well with the results obtained with more sophisticated methods. This is in line with the finding that the second-order correlation energies in $d=3$ and $d=\infty$ do not differ strongly \cite{metzner89}. Finally, results in the thermodynamic limit for the $U$-dependent double occupancy using the numerical linked-cluster expansion show a kink in the double occupation in line with our DMFT  results \cite{khatami_three-dimensional_2016}.

In summary, the DQMC data suggest that, first, a discontinuity in $\partial_{\tilde U} \langle n_0 n_0 \rangle $ sufficiently signals a first-order phase transition at arbitrary small $V$ and, second, the DMFT approximation leads to reasonable results in the three-dimensional case, especially for small interaction strengths.

\section{Strong coupling calculation of CDW phase}
\label{sec:app_strong_coupl}
We calculate the CDW phase with a strong-coupling approach and identify a CDW instability by a negative Fourier component of the Coulomb interaction. For the CDW transition line in the case of the honeycomb lattice we find $U=3V$ and $U\approx1.53V$ for nearest-neighbor and long-range interactions, respectively. For the case of the square lattice we find $U=4V$ and $U\approx 1.61V $ for nearest-neighbor and long-range interactions, respectively. For the interaction strengths of interest ($U/t\sim 8$ and $U/t\sim 2$ for the honeycomb and square lattices, respectively), the strong-coupling result coincides well with more sophisticated calculations for nearest-neighbor interactions \cite{wu_phase_2014,terletska_charge_2017}.

\section{Proof of thermodynamic inequality}
\label{sec:app_thermo}
Let $H=H_0+U\sum_i D_i - \mu \sum_i n_i$, where $D_i$ is the double occupancy of site $i$, $n_i$ is the occupancy of site $i$, and $H_0$ contains all other terms in the Hamiltonian. Then $-U$ and $\mu$ can be interpreted as the Lagrange multipliers fixing the average double occupancy, $D=1/N \sum_i D_i$, and the average particle number, $n=1/N \sum_i n_i$, respectively. The free energy per site of a state with density matrix $\rho$ is given by $f(\rho) = f_0(\rho)+UD-\mu n$, where $f_0(\rho)=1/N\left[E_0(\rho)-TS(\rho)\right]$. The thermodynamic ground state $\rho_0$ minimizes the free energy, so deviations $\delta \rho$ from $\rho_0$ increase the free energy: $\delta f>0$. If we parametrize the density matrix via the double occupancy and the particle number, deviations from the thermodynamic ground state lead to the following changes in the free energy:
\begin{align}
\delta f =& \phantom{+} \frac{\partial f_0}{\partial D} \delta D + U \delta D + \frac{1}{2} \frac{\partial^2 f_0}{\partial D^2} \delta D ^2 \notag \\
& +\frac{\partial f_0}{\partial n} \delta n - \mu \delta n + \frac{1}{2} \frac{\partial^2 f_0}{\partial n^2} \delta n ^2 \notag \\
& + \frac{\partial^2 f_0}{\partial D\partial n} \delta D \delta n. 
\end{align}
The condition that $f$ is at an extremum demands that the first-order terms vanish, i.e., $\partial f_0/\partial D = -U$ and $\partial f_0/\partial n=+\mu$. The second-order term can be written in matrix form as
\begin{align}
\delta f =& \frac{1}{2} 
\begin{pmatrix}\delta D & \delta n\end{pmatrix}
\begin{pmatrix} 
\frac{\partial^2 f_0}{\partial D^2}           &
\frac{\partial^2 f_0}{\partial D\partial n}   \\[0.3em]
\frac{\partial^2 f_0}{\partial D\partial n}   &
\frac{\partial^2 f_0}{\partial n^2}
\end{pmatrix}
\begin{pmatrix}\delta D \\ \delta n\end{pmatrix} \notag \\
=& \frac{1}{2}
\begin{pmatrix}\delta D & \delta n\end{pmatrix}
\begin{pmatrix} 
\frac{-\partial U}{\partial D}           &
\frac{-\partial U}{\partial n}   \\[0.3em]
\frac{\partial \mu}{\partial D}   &
\frac{\partial \mu}{\partial n}
\end{pmatrix}
\begin{pmatrix}\delta D \\ \delta n\end{pmatrix} \label{matrices} 
\end{align}
Now, the condition $\delta f>0$ means that both eigenvalues of this matrix should be positive. Since the matrix is symmetric, this leads to the conditions for thermodynamic equilibrium,
\begin{align}
0 < -\frac{\partial U}{\partial D}, \label{c1}  \\
0 < \frac{\partial \mu}{\partial n}, \label{c2} \\
0 < \frac{\partial U}{\partial n} \frac{\partial \mu}{\partial D} - \frac{\partial U}{\partial D} \frac{\partial \mu}{\partial n}.
\label{c3} 
\end{align}
Equation \eqref{c2} tells us that the compressibility $\kappa=\partial n/\partial \mu$ is positive (at constant double occupancy), and Eq. \eqref{c1} indicates that the double occupancy decreases as a function of $U$ (at constant density). The symmetry of the matrix implies the Maxwell relation $\partial \mu/\partial D = - \partial U/\partial n=A$.

We consider the relation between $\partial D/\partial U$ at constant chemical potential and at constant $n$. We define the implicit function $\mu(U,n)$ to give the chemical potential corresponding to $U$ and $n$, via $n(U,\mu(U,n))=n$. We find
\begin{align}
\left.\frac{\partial n}{\partial U}\right|_\mu + \frac{\partial n}{\partial \mu} \frac{\partial \mu(U,n)}{\partial U}  =0. \label{eq1} 
\end{align}
Using this, we obtain
\begin{align}
\left.\frac{\partial D}{\partial U}\right|_{n} - 
\left.\frac{\partial D}{\partial U}\right|_{\mu} =&
\frac{\partial D}{\partial \mu} \frac{\partial \mu(U,n)}{\partial U} \notag \\
\overset{\eqref{eq1}}{=}& 
- \frac{\partial D}{\partial \mu} \frac{\partial n/\partial U}{\partial n/\partial \mu} \notag  \\
=& \kappa^{-1} A^{-2} \geq 0,
\end{align}
where $\kappa$ is the compressibility and $A$ is the off-diagonal element in Eq.~\eqref{matrices}. The positivity follows since $\kappa$ has to be positive for thermodynamic stability and $A$ appears as a square. Together with Eq.~\eqref{c1} this gives
\begin{align}
0 < \left.-\frac{\partial D}{\partial U}\right|_n \leq \left.-\frac{\partial D}{\partial U}\right|_\mu.
\end{align}

\bibliographystyle{apsrev4-1}
\bibliography{BibliogrGrafeno}

\end{document}